\documentclass[prl,aps,amssymb,superscriptaddress]{revtex4}
\usepackage{amsmath}
\usepackage{amssymb}
\usepackage{amsthm}
\usepackage{amsfonts}
\usepackage{listings}
\usepackage{enumerate}
\usepackage{latexsym}
\usepackage{color}
\usepackage{bm}
\usepackage{hyperref}
\hypersetup{
 pdfnewwindow=true, colorlinks=true,
 linkcolor=blue, anchorcolor=blue,
 citecolor=blue, filecolor=blue,
 menucolor=blue, urlcolor=blue}

\usepackage{psfrag}

\usepackage{bm}
\usepackage{graphicx}
\usepackage{subfigure}

\RequirePackage[normalem]{ulem} 
\RequirePackage{color}\definecolor{RED}{rgb}{1,0,0}\definecolor{BLUE}{rgb}{0,0,1} 

\def\ti{{\it{t}}-LaSbTe}
\def\tci{{\it{o}}-LaSbTe}

\def\ea{{\it et al.}}
\input{epsf}

\begin{document}

\tolerance 10000

\newcommand{\vk}{{\bf k}}

\draft

\title{Layer Construction of Topological Crystalline Insulator LaSbTe  }

\author{Yuting Qian}
\thanks{Those authors contributed equally to this work.}
\affiliation{Beijing National Laboratory for Condensed Matter Physics,
and Institute of Physics, Chinese Academy of Sciences, Beijing 100190, China}
\affiliation{University of Chinese Academy of Sciences, Beijing 100049, China}

\author{Zhiyun Tan}
\thanks{Those authors contributed equally to this work.}
\affiliation{Beijing National Laboratory for Condensed Matter Physics,
and Institute of Physics, Chinese Academy of Sciences, Beijing 100190, China}
\affiliation{School of Physics and Electronic Science, Zunyi Normal University, Zunyi 563006, Guizhou, People's Republic of China}

\author{Tan Zhang}
\affiliation{Beijing National Laboratory for Condensed Matter Physics,
and Institute of Physics, Chinese Academy of Sciences, Beijing 100190, China}
\affiliation{University of Chinese Academy of Sciences, Beijing 100049, China}

\author{Jiacheng Gao}
\affiliation{Beijing National Laboratory for Condensed Matter Physics,
and Institute of Physics, Chinese Academy of Sciences, Beijing 100190, China}
\affiliation{University of Chinese Academy of Sciences, Beijing 100049, China}

\author{Zhijun Wang}
\affiliation{Beijing National Laboratory for Condensed Matter Physics,
and Institute of Physics, Chinese Academy of Sciences, Beijing 100190, China}
\affiliation{University of Chinese Academy of Sciences, Beijing 100049, China}

\author{Zhong Fang}
\affiliation{Beijing National Laboratory for Condensed Matter Physics,
and Institute of Physics, Chinese Academy of Sciences, Beijing 100190, China}
\affiliation{University of Chinese Academy of Sciences, Beijing 100049, China}

\author{Chen Fang}
\email{cfang@iphy.ac.cn}
\affiliation{Beijing National Laboratory for Condensed Matter Physics,
and Institute of Physics, Chinese Academy of Sciences, Beijing 100190, China}
\affiliation{University of Chinese Academy of Sciences, Beijing 100049, China}
\affiliation{Songshan Lake Materials Laboratory, Dongguan, Guangdong 523808, China}
\affiliation{CAS Center for Excellence in Topological Quantum Computation, University of Chinese Academy of Sciences, Beijing 100190, China}

\author{Hongming Weng}
\email{hmweng@iphy.ac.cn}
\affiliation{Beijing National Laboratory for Condensed Matter Physics,
and Institute of Physics, Chinese Academy of Sciences, Beijing 100190, China}
\affiliation{University of Chinese Academy of Sciences, Beijing 100049, China}
\affiliation{Songshan Lake Materials Laboratory, Dongguan, Guangdong 523808, China}
\affiliation{CAS Center for Excellence in Topological Quantum Computation, University of Chinese Academy of Sciences, Beijing 100190, China}

\begin{abstract}
Topological crystalline insulator (TCI) is one of the symmetry-protected topological states. Any TCI can be deformed into a simple product state of several decoupled two-dimensional (2D) topologically nontrivial layers in its lattice respecting its crystalline symmetries called the layer construction (LC) limit. In this work, based on first-principles calculations we have revealed that both tetragonal LaSbTe (\ti) and orthorhombic LaSbTe (\tci) can be interpreted as stacking of 2D topological insulators in each lattice space. The structural phase transition from~\ti~to~\tci~due to soft phonon modes demonstrates how the real space change can lead to the modification of topological states. 
Their symmetry-based indicators and topological invariants have been analyzed based on LC. We propose that LaSbTe is an ideal example demonstrating the LC paradigm, which bridges the crystal structures in real space to the band topology in momentum space.
\end{abstract}

\maketitle
\section{1. Introduction}
 Topological crystalline insulators (TCIs) ~\cite{fu2011topological} are topological states protected by crystalline symmetry as an extension of time-reversal (TR) symmetry protected topological insulators (TIs)~\cite{hasan2010colloquium,qi2011topological,weng2015quantum}. 
Three-dimensional (3D) weak TI protected by lattice translation can be obtained from stacking many copies of 2D TI in the third dimension. 
Heuristically, apart from translation, TCI protected by mirror symmetry has been found to be adiabatically connected to the limit of weakly coupled 2D topological states, and these 2D topological states can be TCIs characterized by mirror Chern number~\cite{kim2015} or 2D Chern insulators~\cite{Fulga2016}. These had inspired researchers to investigate TCIs based on such dimensional reduction argument, named as "layer construction" (LC) scheme now. 
It has been demonstrated~\cite{huang2017,hsong2017,Songzdarxiv1,Songzdarxiv2} that all crystalline point group symmetry-protected topological phases can be produced by stacking lower-dimensional blocks of topological states and the classification can be applied to both interacting bosons~\cite{else2018} and free fermions~\cite{prx_ashvin,song2018quantitative}.
This approach provides a clear understanding of the physical nature of the crystalline symmetry-protected topological phase. 

Song ~\ea~\cite{song2018quantitative} have exploited the LC scheme in very detail. They derived exhaustive mappings from symmetry data to topological invariants for arbitrary space groups in the presence of TR symmetry, providing an effective approach to define and search for new TCIs~\cite{Zhang2018,Vergniory2019,wanxg2019} when combined with topological quantum chemistry (TQC)~\cite{tqc2017,nc_ashvin,Kruthoff2017} and symmetry-based indicators (SBI).
LC decorates the lattice with 2D insulating layers with nontrivial $\mathbb Z_2$ number or mirror Chern number, while atomic insulators are constituted from proper atomic orbitals at Wyckoff positions. But the correspondence between an LC and the resulting TCI is not as intuitive as the atomic picture.
Recently, A. Matsugatani~\ea~connected the high-order TIs (HOTIs) to lower-dimensional TIs using a tight-binding model to capture the essence of HOTIs~\cite{haruki2018}. However, a realistic TCI material with layered crystal structure has yet to be demonstrated with exact one-to-one correspondence between the stacking of lower-dimensional topological phases in real space and the band topology.

In this paper, we introduce LaSbTe with two distinct crystal structures: tetragonal LaSbTe (\ti) (a member of $WHM$ ($W$ = Zr, Hf, or La, $H$ =Si, Ge, Sn, or Sb, and $M$= O, S, Se, or Te) family~\cite{xu2015two}) and orthorhombic LaSbTe (\tci). When spin-orbit coupling (SOC) is taken into consideration, ~\ti~is a weak TI with one 2D TI in its unit cell, while~\tci~is a TCI with two 2D TI layers per unit cell due to doubling of~\ti~cell along the stacking direction during phase transition. Thus, \tci~can be viewed as an LC of two~\ti~with proper lattice distortion, where each~\ti~represents a layer of an elementary LC (eLC). To the best of our knowledge, this is the first demonstration of the LC scheme, in other words, mapping the crystal structures in real space to the band topology in momentum space and paves the way for understanding the physical essence of TCI.

\begin{figure*}[!htb]
\centering
\includegraphics[width=12 cm]{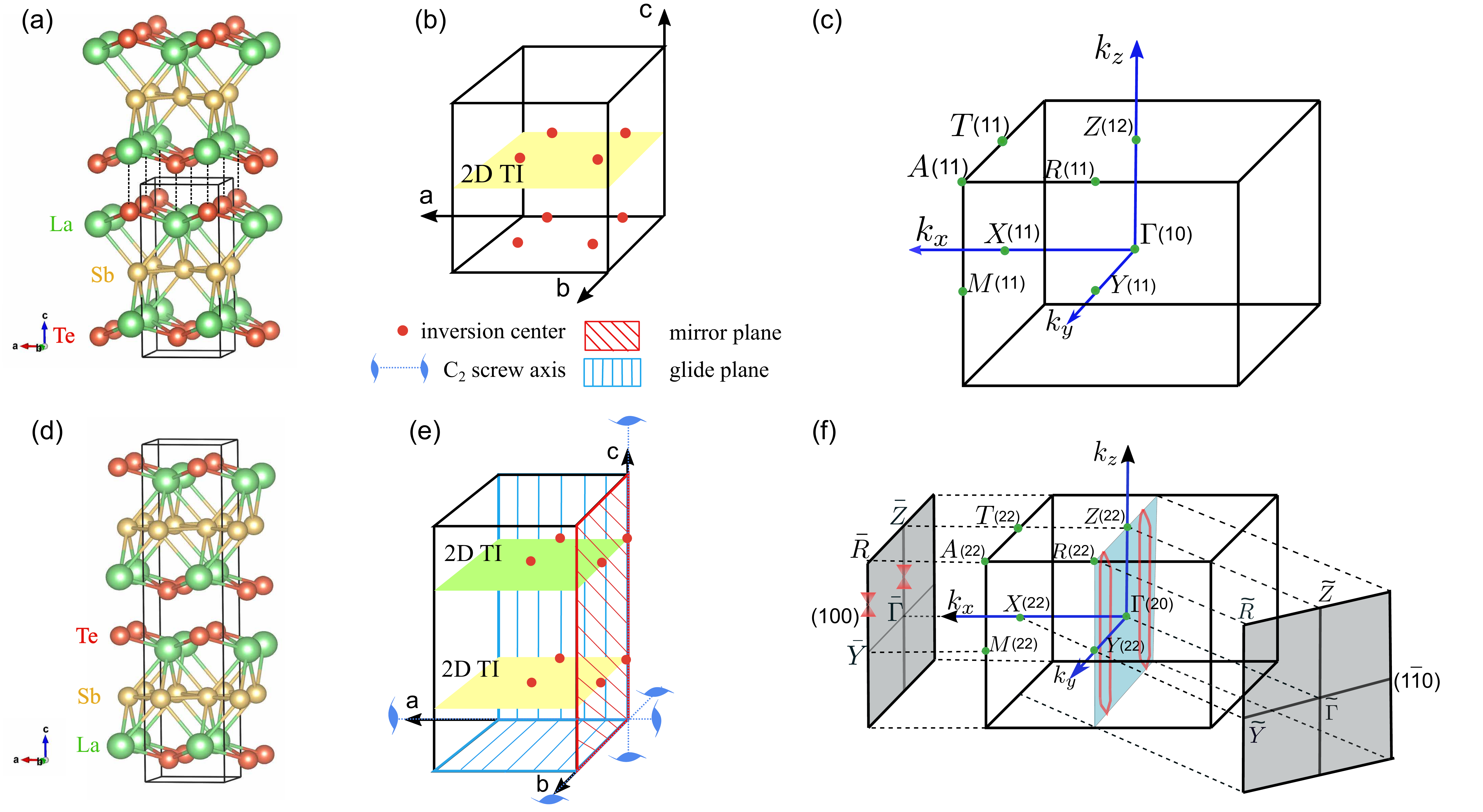}
\caption{(Color online)
(a) Crystal structure of~\ti. (b) Layer Construction (LC) of a weak TI as the stacking of 2D TI (yellow plane) in lattice space corresponding to~\ti. The red dots are inversion centers.
(c) Bulk Brillouin zone (BZ) and high symmetrical crystal momenta for~\ti. The number in parentheses near high symmetrical momenta is the number of occupied Kramers pair bands with negative parity eigenvalues. (d)-(f) are for~\tci. In (e), the mirror plane, glide plane and screw axis are also shown. In (f) the blue plane indicates the mirror plane and the red lines inside of it represent the nodal rings calculated without spin-orbit coupling (SOC). The dark gray planes are the projected surface BZ for (100) and (1$\bar 1$0) surfaces of~\tci,~where the red hourglass shapes represent the hourglass-like surface states along the two paths. 
}
\label{fig:1}
\end{figure*}

\section{2. Methodology}
The Vienna {\it ab initio} simulation package (VASP)~\cite{vasp} with the projector augmented wave method~\cite{paw1,paw2} based on density functional theory was employed for the first-principles calculations. The generalized gradient approximation (GGA) of Perdew-Burke-Ernzerhof type~\cite{pbe} was used for the exchange-correlation potential. The cut-off energy for plane wave expansion was set to 300 eV. The Brillouin zone (BZ) was sampled as grids of $8\times 8\times 4$ for \ti~and $12\times 12\times 3$ for \tci~in the self-consistent process, respectively.  A width of 0.02 eV was adopted in the Gaussian smearing method for Fermi level determination. We employed the experimental lattice parameters~\cite{charvillat1977cristallochimie,dimasi1996stability} and then fully relaxed the structure until the forces on all atoms were smaller than 0.01 eV/\AA. The band structures were calculated with and without considering of SOC. 
Because of the weak coupling between quintuple layers (QLs), different forms of van der Waals (vdW) functional, including optPBE-vdW, optB88-vdW and optB86b-vdW were adopted to check the effect of vdW interactions on the topology~\cite{Klime2010,Klime2011,vdw2007}.
The WannierTools~\cite{WU2017} was employed  to compute the surface states based on the maximally localized Wannier function (MLWF) constructed by Wannier90~\cite{2012maximally}. The phonon spectrum was computed by the Phonopy program based on the density functional perturbation theory (DFPT)~\cite{togo2015first}. The Wilson-loop technique was used to compute TR $\mathbb Z_2$ in a single-layer calculation.

\section{3. Crystal structure}

\begin{figure}[!htb]
\centering
\includegraphics[width=10cm]{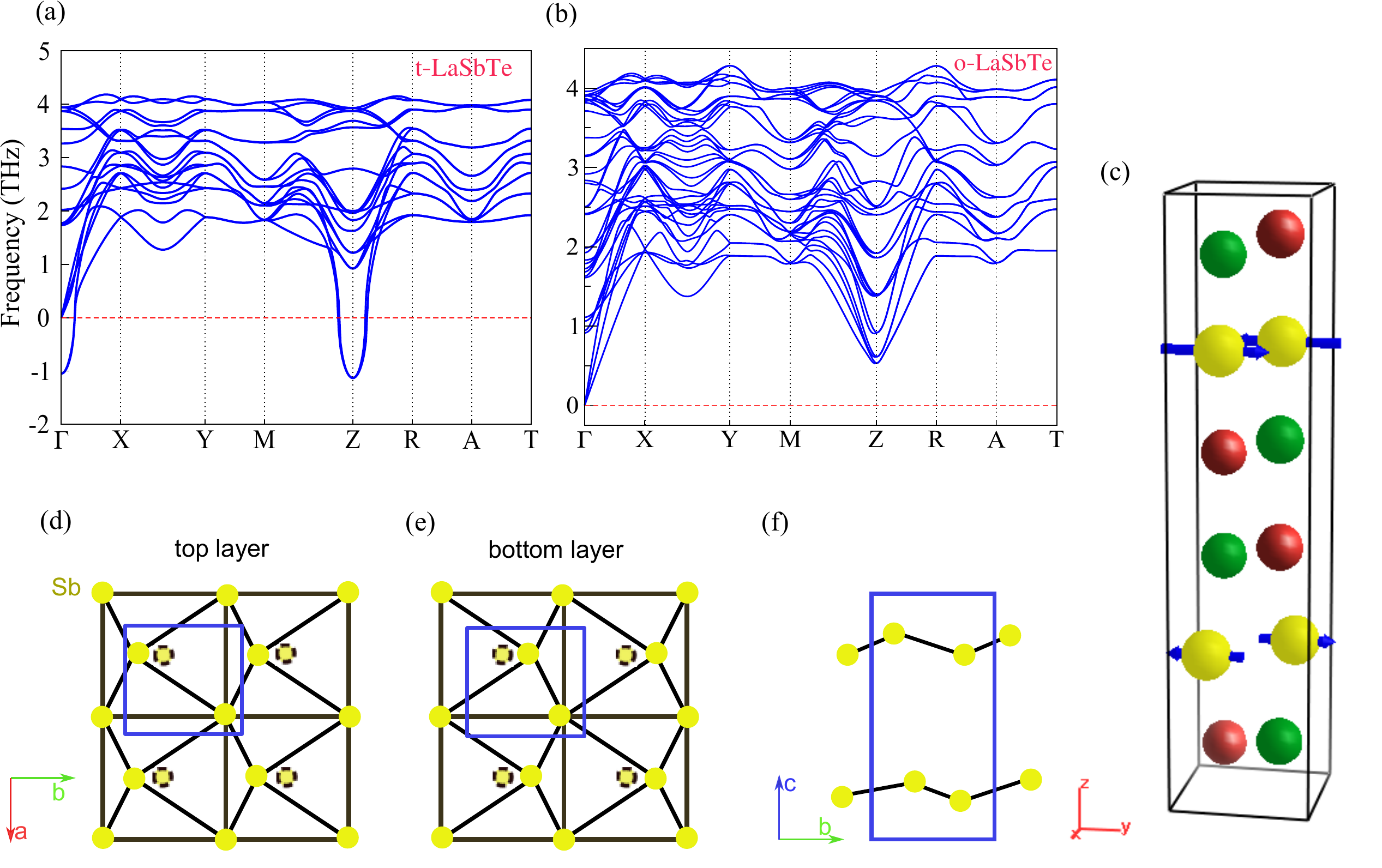}
\caption{(Color online)
Calculated phonon spectra for (a)~\ti~ and (b)~\tci~, respectively. (c) Two unit cells of~\ti~along the c-axis with arrows on Sb atoms denoting the displacement in one of the soft phonon modes at $Z$. The view from $c$-direction of the (d) top and (e) bottom Sb layer for a $4 \times 4 \times 1$~\tci~supercell (black box). The blue box represents the actual selection of the primitive cell. The solid yellow circle represents Sb atoms in~\tci~due to the distortion from those in ~\ti~(dotted yellow circles) consistent with the soft modes in (c).  (f) The side view of~\tci~from $a$-axis shows the zigzag chain like Sb atoms along the $b$-axis.
}
\label{fig:2}
\end{figure}

Two different crystal structures have been reported for LaSbTe. One is~\ti~with tetragonal phase and space group (SG.) $P4/nmm$ (No. 129). The relaxed lattice constants are 
$a=b= 4.421$~\AA~and $c= 9.659$~\AA~that are comparable with experimental values $a=b=4.44$~\AA~and $c= 9.47$~\AA. La and Te are at Wyckoff position $2c$ $(0.5, 0.0, z)$ with $z=0.7785$ and $0.1282$, respectively. Sb is at $2b$ $(0.0, 0.0, 0.5)$. \ti~is one of the $WHM$ family materials,~\cite{xu2015two} which comprise stacked quintuple layers along the $c$-axis, where QL is composed of five square nets in a sequence of [Te-La-Sb$_{2}$-La-Te]~\cite{wang1995main}.
QLs are bonded with each other along the $c$-direction, as indicated by the dotted lines in Fig.~\ref{fig:1} (a) and they are suggested to be weaker than the bonds among square nets in QL. 

The other one is~\tci~in orthorhombic phase with SG. $Pmcn$ (No. 62) with relaxed lattice constants $a=4.422$~\AA, $b=4.433$~\AA, and $c=19.348$~\AA. The experimental lattice constants are $a=4.378$~\AA, $b=4.403$~\AA, and $c=19.242$~\AA. La, Te and Sb are at Wyckoff position $4c$ $(0.2500,0.2623,0.6106)$, $(0.2500,0.2614, 0.4360)$, and $(0.2500,0.7283,0.2491)$, respectively. Compared with~\ti, two QLs take place in one unit cell of~\tci~as the dimerization of QLs since the square-sheet layer of Sb in each QL is distorted, while Sb sheets in neighboring QLs distort in the opposite direction. 
\tci~has the inversion symmetry, a mirror plane ($m^{100}_{\frac{1}{2}00 }$), two glide planes ($g^{010}_{0\frac{1}{2}\frac{1}{2}}$, $g^{001}_{\frac{1}{2}\frac{1}{2}0}$) and three two-fold screw axes ($2^{100}_{ \frac{1}{2}0\frac{1}{2}}$, $2^{010}_{0 \frac{1}{2}0}$, $2^{001}_{\frac{1}{2}\frac{1}{2}\frac{1}{2}}$).

The structural difference in these two phases can be understood by studying their phonon spectra, which are calculated and shown in Fig.~\ref{fig:2}(a) and (b). The phonon spectrum of~\tci~indicates it is the stable phase, while that of~\ti~has soft modes with imaginary frequencies at $\Gamma$ and $Z$. The soft modes at $Z$ (0, 0, 0.5) suggest that the doubling of the $c$-axis would lead to a stable crystal structure. These soft modes mainly consist of Sb vibration modes as shown in Fig.~\ref{fig:2}(c), where Sb atoms in neighboring QLs shift in the opposite direction and lead to a rectangular in-plane lattice and a zigzag chain-like structure, as schematically depicted from both top and side views in Fig.~\ref{fig:2}(d-f). Such kind of lattice distortion corresponds to a structural transition from~\ti~to~\tci.
The soft modes at $\Gamma$ are from optical phonon branches, which would result in a phase transition to SG. 11. Nonetheless, the transition to~\tci~with SG. 62 takes place, while the phase of SG. 11 has not been reported.

\section{4. Electronic band structures}
\begin{figure}[!htb]
\centering
\includegraphics[width=11 cm]{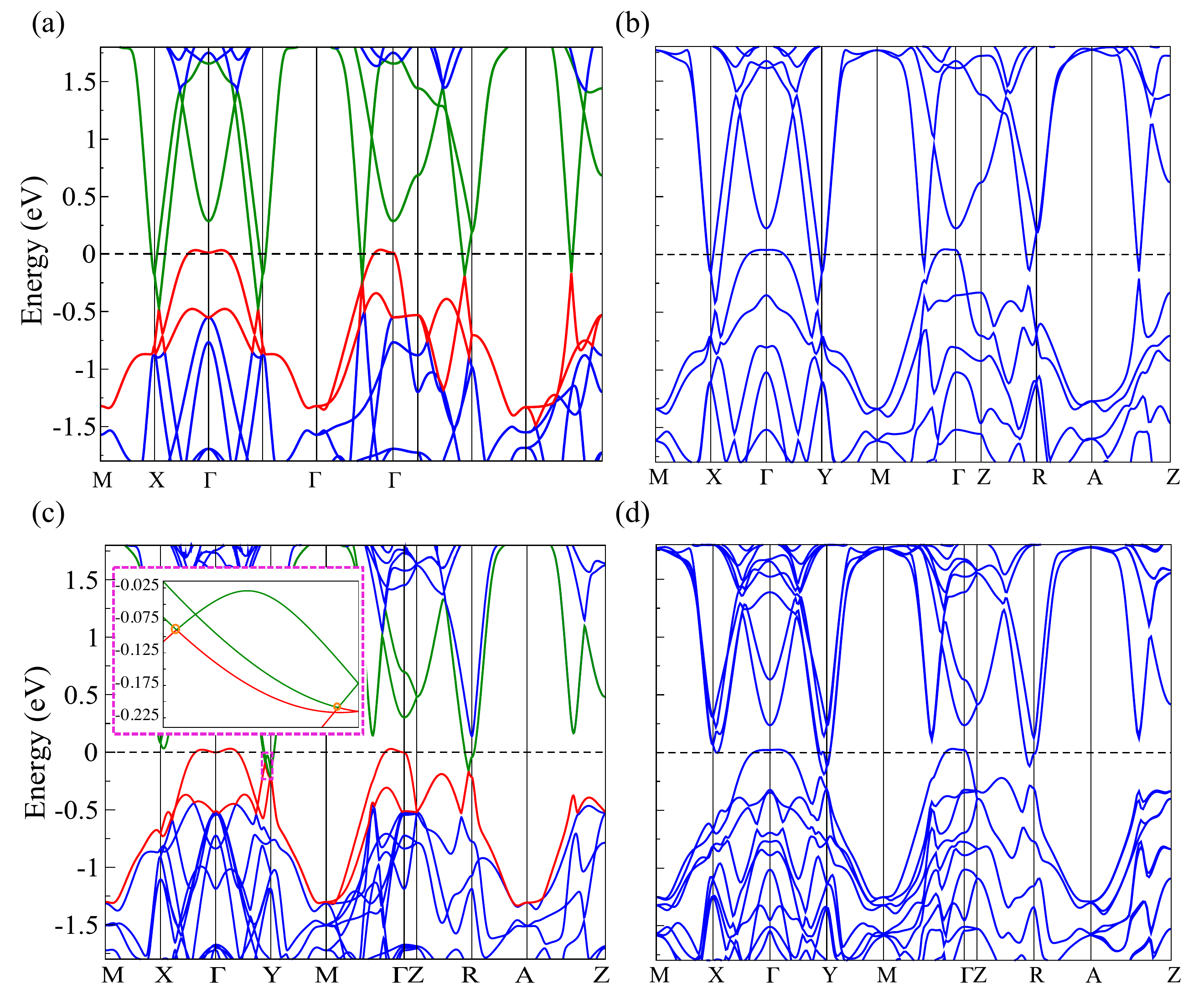}
\caption{(Color online)
Calculated band structures along the high-symmetry lines within GGA without ((a),(c)) and with ((b),(d)) SOC for \ti~((a),(b)) and~\tci~((c),(d)). The inset in (c) shows the band inversion feature along $\Gamma-Y$ without SOC for~\tci.
The two highest valance bands and lowest conduction bands are marked in red and green, respectively.
}
\label{fig:3}
\end{figure}

The band structures for~\ti~and~\tci~without and with SOC are shown in Fig.~\ref{fig:3}. As a family of $WHM$ compounds, \ti~has nodal lines when SOC is neglected~\cite{xu2015two, lou2016emergence} since there are band crossings along $\Gamma-X (Y)$, $\Gamma-M$ in $k_z=0$ plane and $Z-R$, $Z-A$ in $k_z=\pi$ plane. All the nodal lines are gapped when SOC is further included and \ti~becomes a weak TI as a simple stacking of QLs (i.e., 2D TIs)~\cite{xu2015two}. There are also crossing points in the band structure of~\tci~without SOC, which are parts of the nodal lines in the mirror plane $m^{100}_{\frac{1}{2}00 }$ as schematically shown in Fig.~\ref{fig:1}. Again, SOC opens the band gap along the nodal lines and~\tci~becomes a TCI.
Neither t-LaSbTe nor o-LaSbTe has a global band gap to be a real insulator; however, this might be improved by replacing the component atoms with heavier elements to enhance SOC splitting, or by proper lattice strain and pressure effect.

According to the theory of SBI~\cite{nc_ashvin,wanxg2019}, both~\ti~and~\tci~have inversion symmetry, and when SOC is considered, their symmetry indicators are $\mathbb Z_{2,2,2,4}$ , which can be easily calculated through Fu-Kane like formula~\cite{song2018quantitative, prx_ashvin}. The three weak $\mathbb Z_2$  indices, denoted as $z_{2w,i=1,2,3}$, are obtained through
\begin{equation}
\label{eq:1}
\begin{aligned}
z_{2w,i=1,2,3}\equiv&\sum_{\rm {\bold K}\in TRIM \;at \;\{k_i=\pi\}} n^{-}_{\bold K} \; \mathrm {mod} \;{2}
\end{aligned} .
\end{equation}

The $\mathbb Z_4$ indicator, denoted as $z_4$ is from:
\begin{equation}
\label{eq:2}
\begin{aligned}
 z_4  \equiv \sum_{\rm \bold K \in TRIM} \frac {n^{-}_{\bold K} -n^{+}_{\bold K}}{2}  \; \mathrm {mod} \;{4}
 \end{aligned} ,
\end{equation}
where the $n^{+}_{\bm K}$ ($n^{-}_{\bm K}$) is the number of occupied  Kramers pairs with $+$ $(-)$ parity at time-reversal invariant momentum (TRIM) $\bm K$.
The number of occupied Kramers pairs with negative parity at eight TRIMs in the BZ has been marked in Fig.~\ref{fig:1}. Therefore, according to eq.~\ref{eq:1} and eq.~\ref{eq:2}, we can get~\ti~is a weak TI with $\mathbb Z_{2,2,2,4}$ being (0010) and~\tci~is a TCI of (0002). 
Due to weak coupling between QLs, we adopt different vdW functionals to check the impact of vdW force on the topology~\cite{PRB97}. After fully relaxed, we found optB86b-vdW functional gives a smaller $c$ lattice parameter than PBE functional, 9.58 \AA~for~\ti~ and 19.18 \AA~for~\tci, respectively. The topological properties are the same as our previous calculations. Therefore, we use the PBE functional for all subsequent calculations.
It is known that each single layer (QL) of~\ti~is a 2D TI~\cite{xu2015two}. After the phase transition, the distorted QL is still a 2D TI.[More details can be found in the Appendix A1.] The two QLs in one unit cell of~\tci~are weakly coupled as a simple stacking.
These differences, as well as their corresponding topological invariants and topological properties can be further understood through the LC scheme, which relates the real space crystal structure with the band topology in momentum space.

\section{5. LC and Topological Invariants}
Intuitively,~\ti~can be looked as being constructed by putting a layer of 2D TI on the planes with Miller indices of (001; 1/2). Applying all the symmetrical operations of SG. 129 on this layer gives out the same Miller planes. Therefore, eLC of (001; 1/2) is shown in Fig.~\ref{fig:1}(b). It intersects $a$-axis and $b$-axis zero times and $c$-axis once per unit cell, leading to the three weak topological invariants $z_{2w, i=1, 2, 3}$ being 0, 0 and 1, respectively. The eLC of (001; 1/2) occupies only four inversion centers out of eight, which means the topological invariant $\delta_i$ protected by inversion symmetry for this eLC is convention-dependent and $z_4=0$, $\delta_i=0$. This eLC doesn't occupy any rotation or screw axis, nor any mirror or glide plane. Further, it doesn't lead to nonzero stacking invariant for any screw and glide symmetry. These mean the topological invariant due to rotation, screw, mirror or glide symmetries are all zero, and thus~\ti~is a weak TI instead of TCI.

The structural phase transition from~\ti~to~\tci~has been revealed by the soft modes at $Z$ of~\ti, leading to doubling of the unit cell along the $c$-axis and shifts of atoms. Therefore, two 2D TIs decorating the lattice of~\tci~in the Miller planes of (001; 1/4) and (001; 3/4) constitute the eLC of (001; 1/4) or (001;3/4) within the lattice of SG. 62. In the convention of choosing unit cell in this work, these two layers intersect all the three lattice vectors even times and $z_{2w, i=1, 2, 3}=0$. All the inversion centers, as marked in Fig.~\ref{fig:1}(e), are occupied by these two 2D TI layers; thus, $\delta_i=1$ and $z_4=2$. Again, these two layers are not in any mirror or glide plane, nor do they pass through any rotation axis. However, they contribute nonzero glide-stacking-number for glide symmetry $g^{010}_{0\frac{1}{2}\frac{1}{2}}$ and nonzero screw-stacking-number for screw symmetries $2^{100}_{ \frac{1}{2}0\frac{1}{2}}$ and $2^{001}_{\frac{1}{2}\frac{1}{2}\frac{1}{2}}$. The corresponding topological hourglass invariant $\delta_h$ and screw invariant $\delta_s$ equal to one. By this approach, the topological state protected by TR and crystalline symmetries of~\tci~has been fully determined. It is a TCI with symmetry indicators of $\mathbb Z_{2,2,2,4}=(0002)$ and topological invariants of $\delta_i=1$, $\delta_{h}^{(010)}=1$ and $\delta_{s}^{(100)}=\delta_{s}^{(001)}=1$. More details about the LC of~\tci~are provided in the Appendix A2.


\begin{figure*}[!htb]
\centering
\includegraphics[width=18 cm]{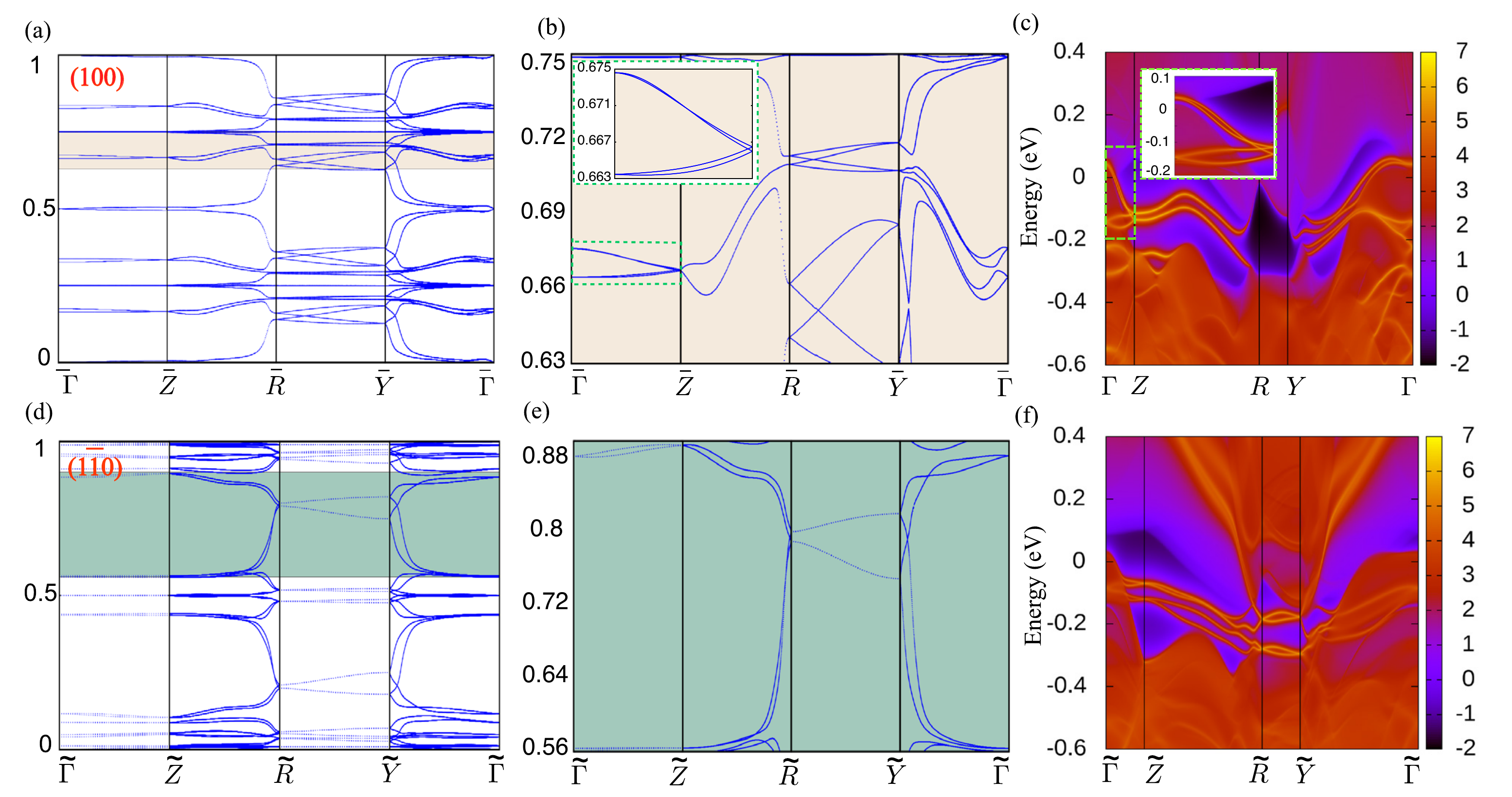}
\caption{(Color online)
The spectra of Wilson loop and SS band structures on surface BZ of (100) (a-c) and (1$\bar 1$0) (d-f) surfaces, respectively. (b, e) The zoomed in image of the shadowed part in (a) and (d), respectively. Hourglass-like SSs in (b, c) but trivial SSs with full gap in (e, f) can be observed. The insets in (b) and (c) indicate the Hourglass-like SSs.
}
\label{fig:4}
\end{figure*}

\section{6. Topological Surface States from Topological Invariants}
The nonzero hourglass invariant $\delta_h$ suggests that nontrivial hourglass surface state (SS) exists on the surface that preserves corresponding glide mirror symmetry~\cite{wang2016hourglass,bernevigprx}.
It is known that the Wilson-loop spectrum defining the bundle of occupied states over the BZ is isomorphic to the spectrum of SS~\cite{fidkowski2011model}. The hourglass invariant can be determined by calculating either the spectrum of Wilson loop based on bulk Hamiltonian, or the SS on the corresponding surface based on surface Green's function method with truncated surface as the boundary condition~\cite{WU2017,Sancho_1985}.
For the hourglass invariant $\delta_{h}^{(010)}=1$, it protects the nontrivial hourglass SS along the paths $\bar\Gamma-\bar Z$ and $\bar R-\bar Y$ on (100)-surface as shown in Fig.~\ref{fig:1}(f) since the glide plane $g^{010}_{0\frac{1}{2}\frac{1}{2}}$ is preserved and projected to them. 
The Wilson-loop spectra in Fig.~\ref{fig:4} (a) and (b) clearly show that there are hourglass-like SSs along $\bar\Gamma-\bar Z$ and $\bar R-\bar Y$. The connecting curves $\bar Z-\bar R$ between these two hourglasses with a generalized zigzag pattern exhibit a spectral flow~\cite{bernevigprx}. In the SS calculation within Green's function method, the hourglass-like SS along $\bar\Gamma-\bar Z$ can be clearly reproduced in Fig.~\ref{fig:4}(c). However, the one along $\bar R-\bar Y$ is hardly seen since it is mixed with the gapless bulk states.
Similarly, $\delta_{h}^{(001)}$ is trivial and there should be no topological SS when projected along $\tilde Z-\tilde R$ and $\tilde Y-\tilde\Gamma$ on (1$\bar1$0)-surface to maintain the glide symmetry $g^{001}_{\frac{1}{2} \frac{1}{2} 0 }$.
The hourglass-like spectra along $\tilde Z-\tilde R$ and $\tilde Y-\tilde\Gamma$ shown in Fig.~\ref{fig:4}(d) and (e) are trivial since the connection between them shows a spectrally isolated quadruplets with a full gap when viewed along $\tilde Z\tilde R\tilde Y\tilde\Gamma$ in whole surface BZ and is consistent with the trivial topological SS band structure in Fig.~\ref{fig:4}(f).

The nontrivial screw invariant $\delta_s$ can lead to protected hinge states, which have been carefully discussed in several studies~\cite{wang2016hourglass,Fang2017,zhangt2019} and will not be presented here for simplicity.

\section{7. Conclusion}
Based on the first-principles calculations, we conclude that a TCI phase~\tci~ with hourglass SS and hinge state can be viewed as stacking of two~\ti~due to structural phase transition. This dimensional reduction method can effectively capture the physics of TCI. 
To the best of our knowledge, this is the first demonstration of a paradigm that visually maps the LC in real space to the band topology in momentum space for experimentally available material. For example, Ba$_3$Cd$_2$As$_4$ \cite{zhangt2019} can also be considered as an LC: $(001;0){\otimes} (001;1/2)$, but it cannot give an image that directly corresponds to the crystal structure in real space. Our paradigmatic demonstration offers an excellent platform to  understand the underlying physics of TCIs and the relationship between TIs and TCIs.

\noindent \textbf{Acknowledgments}
We acknowledge helpful discussions with Yi Jiang and Zhida Song.
This work was supported by the National Natural Science Foundation of China under Grant Nos. 11504117, 11674369,11925408, 11921004 and 11974395.
Zhiyun Tan acknowledges the Foundation of Guizhou Science and Technology Department under Grant No. QKH-LHZ[2017]7091.
Zhijun Wang acknowledges support from the National Thousand-Young-Talents Program, the CAS Pioneer Hundred Talents Program, and the National Natural Science Foundation of China.
 Hongming Weng acknowledges support from the National Key Research and Development Program of China (Grant Nos. 2016YFA0300600, 2016YFA0302400, and 2018YFA0305700), the K. C. Wong Education Foundation (GJTD-2018-01).



\appendix
\section{APPENDIX}
\subsection*{\label{sec:level1} A1. Single-layer calculations of~\tci~ }
\begin{figure*}[!htb]
\centering
\includegraphics[width=14cm]{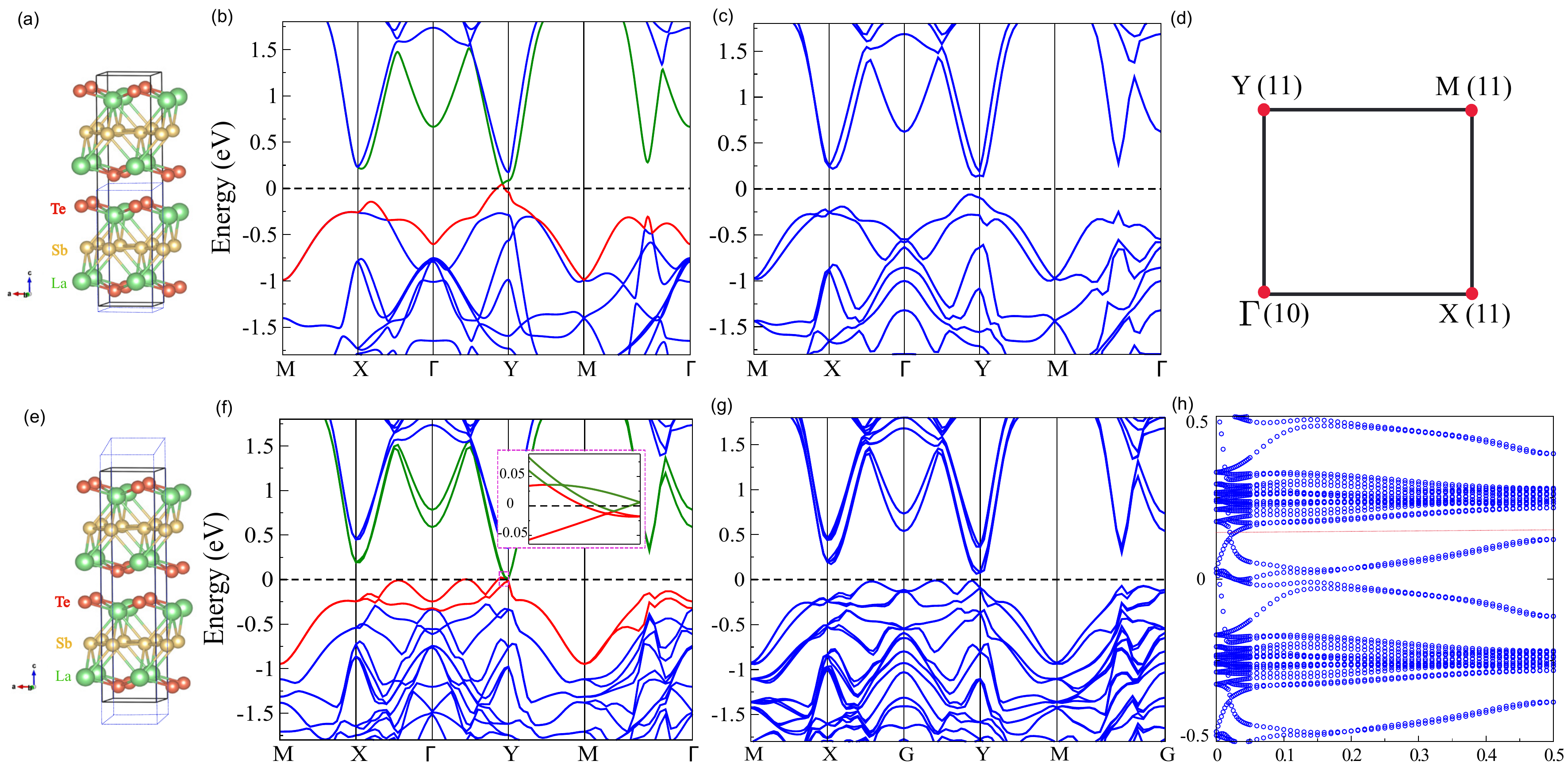}
\caption{(Color online)
Single-layer calculations for one QL and two distorted QLs  of~\tci~. The conresponding primitive cell is represented by a blue box in (a) and (e),respectively. Calculated band structure along the high-symmetry lines within GGA without ((b),(f)) and with ((c),(g)) SOC for one QL and two QLs of~\tci, respectively. The inset in (f) shows the band inversion feature along $\Gamma-Y$. (d) The 2D BZ for these two single-layer structures. The numbers in parentheses near high symmetrical momenta are the numbers of occupied Kramer pair bands with negative parity eigenvalues for one QL of ~\tci~to determine the time-reversal $\mathbb Z_2$. (h) The calculated time-reversal $\mathbb Z_2$ in $k_z=0$ with Wilson-loop method for two QLs of~\tci~, where the red dashed line is a reference line. 
}
\label{figs:1}
\end{figure*}

 In this section, we determine whether one QL of~\tci~ is a 2D TI and two QLs of~\tci~is a trivial insulator. First, we construct a corresponding single-layer structure for these two cases, as shown in Fig.~\ref{figs:1}(a) and (e). Then, we compute the band structures without and with considering SOC. Without SOC, there are one band inversion and double band inversions along $\Gamma-Y$ for one QL and two QLs, respectively. When SOC is considered, these two cases are fully gapped so that the TR $\mathbb Z_2$ can be well defined.  Since one QL keeps inversion symmetry, $\mathbb Z_2$ can be simply derived by Fu-Kane formula. According to eq.1, we obtain $\mathbb Z_2=1$ for one QL. However, for two QLs, inversion symmetry is broken. Therefore, the Wilson-loop method is employed to calculate the Wannier-center flow. $\mathbb Z_2$ equals zero because the number of crossings between the Wilson-loop bands and reference line is even.
By calculating $\mathbb Z_2$ index, we confirm each QL of~\tci~ is a well-defined 2D TI and stacking two weakly coupled 2D TIs can generate a normal insulator with trivial $\mathbb Z_2$  index.

\subsection{\label{sec:level2} A2. LC of~\tci~ }
The LC method has been introduced to map the symmetry-based indicator and topological invariants~\cite{song2018quantitative}. For~\ti~with SG.$P4/nmm$, our convention is the same as in Ref~\cite{song2018quantitative}, so we can directly use the eLC of SG.129 as provided in Supplementary Table 5 of Ref~\cite{song2018quantitative}.  However, for~\tci~with SG.$Pmcn$ in this work, the convention is different from that in Ref~\cite{song2018quantitative}. Then, we derive the eLC of SG.$Pmcn$ and tabulate it in  Table~\ref{table:1}. The second eLC (001;1/4) in  Table~\ref{table:1} represents the eLC of~\tci.

\begin{table*}[!htb] 
\footnotesize
\caption{Elementary Layer Construction for Space Group $Pmcn$}
\label{table:1}
\tabcolsep 16pt 
\begin{tabular*}{\textwidth}{c|c|c|c|c|c|c|c|c|c}
 \hline
  \hline
 $(hkl;d)$ & $\mathbb Z_{2,2,2,4}$ & weak & $m^{100}_{(2)}$ &  $g^{010}_{0 \frac{1}{2}\frac{1}{2} }$&$g^{001}_{\frac{1}{2} \frac{1}{2} 0 }$& $i$& $2^{100}_{1 }$ & $2^{010}_{1}$ & $2^{001}_{ 1}$ \\
\hline
 001;0 & 0000 & 000 & 00 &  1 & 1 & 0 & 0 & 1&1 \\
 \hline
001;$\frac{1}{4}$& 0002& 000 & 00 &   1 & 0 &  1 & 1&  0& 1 \\
 \hline
 100;$\frac{1}{4}$ & 0000& 000 & 20 &  0 & 1 & 0 & 1 & 0&1 \\
 \hline
 \hline
\end{tabular*}
\end{table*}


\end{document}